# Element Doping Enhanced Charge-to-Spin Conversion Efficiency in Amorphous PtSn$_4$ Dirac Semimetal

Jinming Liu[a,†], Yihong Fan[a,†], Delin Zhang[a,†,*], Onri J. Benally[a], Lakhan Bainsla[a], Thomas Peterson[b], and Jian-Ping Wang[a]

[a] Department of Electrical and Computer Engineering, University of Minnesota, Minneapolis, MN 55455, United States

[b] School of Physics and Astronomy, University of Minnesota, Minneapolis, MN 55455, United States

**Abstract**

Topological semimetals (TSs) are promising candidates for low-power spin-orbit torque (SOT) devices due to their large charge-to-spin conversion efficiency. Here, we investigated the charge-to-spin conversion efficiency of amorphous $PtSn_4$ (5 nm)/CoFeB (2.5-12.5 nm) layered structures prepared by a magnetron sputtering method at room temperature. The charge-to-spin ratio of $PtSn_4$/CoFeB bilayers was 0.08, characterized by a spin torque ferromagnetic resonance (ST-FMR) technique. This ratio can further increase to 0.14 by inducing dopants, like Al and CoSi, into $PtSn_4$. The dopants can also decrease (Al doping) or increase (CoSi doping) the resistivity of $PtSn_4$. The work proposed a way to enhance the spin-orbit coupling (SOC) in amorphous TSs with dopants.

1. **Introduction**

Spin-orbit torque (SOT) has been attracting a significant research interest after the discovery of a giant spin Hall Effect (SHE) in heavy metals with large spin-orbit interactions.[1–3] A charge current $J_c$ flowing through a material with a strong SOT coupling can generate a transverse spin current $J_s$ that helps switch the magnetization of the adjacent ferromagnet (FM) layer.[4] It provides an alternative approach to manipulate spin currents besides the spin transfer torque (STT) based magnetic devices that require an FM 'polarizer' layer to generate a spin current and switch the magnetic layer.[5] Thus, the spin channel used to obtain spin currents does not have to be FM anymore, and no large charge current would be required to apply to the FM layers either. Additionally, the SOT effect can switch an adjacent FM layer more efficiently than that of STT effect due to the larger charge-to-spin conversion efficiency.[6] Materials with large SOT effects are highly demanded and intensively studied.

Recently, besides heavy metals like Pt, topological materials such as Weyl and Dirac topological semimetals (TSs) are attracting increasing attentions due to their large spin torque efficiency $J_s/J_c$ that could be larger than 100 %.[6,7] The spin torque generated in TSs is mainly due to the spin-momentum locking of the spin of electrons, bulk spin Hall effect, and interfacial Rashba effect. The research based on topological materials primarily focus on crystalline materials grown by molecular beam epitaxial approach or by mechanical exfoliating some flakes from bulk single crystals.[8,9] $PtSn_4$ single crystal is one of the example, which has been successfully synthesized by many researchers and demonstrated that $PtSn_4$ has Dirac node arcs forming closed loops in the momentum space.[10–12] Large angular dependent magnetoresistance (AMR) and giant planar Hall effect are reported in single crystal of $PtSn_4$, which makes it a potential candidate for SOT based spintronic devices.[13,14]

A recent theoretical work demonstrated a topological phase in 3D amorphous metals showing a similar and even larger spin Hall conductivity over their crystalline counterparts.[15] Additionally, the states of spin polarized edges can keep the topological protection in an amorphous system.[16] For example, the spin swapping effect of amorphous $WTe_{2-x}$ was reported to contribute spin-orbit scatterings originated from the amorphous structure.[17] Amorphous TSs show a similar charge-to-spin conversion efficiency to the crystalline ones such as $Bi_xSe_{1-x}$.[6,14] Due to the quantum confinement effect originated from the reduction of grain sizes, a larger spin torque efficiency of $Bi_xSe_{1-x}$ has been reported on thin films with nanocrystalline structure compared with a single crystal film prepared by a molecular beam expitaxy approach.[7,18] Additionally, amorphous thin films are more homogenous and not strongly affected by the size and dimension of SOT devices, which could ensure a better scalability. However, few experimental works on amorphous TSs have been reported yet.

The SHE is essential in the spintronics, and it is crucial to seek methods to enhance such effects. However, the electronic structures of a specific material determine the intrinsic mechanisms for SHE. Some extrinsic factors like skewing scattering can also contribute to SHE because of the electron scattering to enhance the spin-orbit coupling (SOC). The intrinsic properties of SHE is difficult to manipulate. Thus, many ways have been proposed to enhance the SHE extrinsically, such as inducing electron scattering centers in a spin channel to enhance the SOC. For example, Azevedo et al. reported the increase of spin Hall angle in Pt/yttrium iron garnet (YIG) layered structures by inducing Ag particles in the midplane of the Pt layer.[19] Zhu et al. also demonstrated the giant spin Hall angle up to 0.73 by dispersing MgO impurities into the Pt layer.[20] Thus, we use Al and CoSi as dopants in the amorphous $PtSn_4$ layer to induce the electron scattering at interfaces around the dopants, which may increase the SOC and spin Hall angle accordingly.

In this work, we use a magnetron sputtering approach to synthesize amorphous $PtSn_4$ thin films to investigate the potential application on SOT spintronic devices. Thin films were deposited at room temperate with an amorphous structure. CoFeB layer with different thicknesses were used as the FM layer adjacent to $PtSn_4$ to characterize the charge-to-spin conversion efficiency. In addition, dopants, like Al and CoSi, were also induced into PtSn$_4$ thin films to tune the SOC and further enhance the change to spin efficiency. Al doping can also decrease the resistivity of PtSn$_4$, which reduces the shunting currents in the adjacent FM layer and then lower the critical current required to switch the FM layer. CoSi doping increases the resistivity that enhance the scattering of carriers and the charge to spin efficiency.

## 2. Methods

All the samples studied in this work were deposited on single crystal MgO (001) substrates by a magnetron sputtering method under an ultrahigh vacuum (base pressure $< 5.0 \times 10^{-8}$ Torr). Ar gas pressure during the sputtering was 2.0 mTorr for all the depositions. The bilayer structure with a spin channel (PtSn$_4$, PtSn$_4$:Al, and PtSn$_4$:CoSi) and a FM adjacent layer (CoFeB) were synthesized at room temperature, and a 2 nm Al layer was used as a capping layer on top of the CoFeB layer. The atomic ratio between the dopants, like Al or CoSi, and PtSn$_4$ was 1:4. The dopants were induced using a co-sputtering technique based on the deposition rates of PtSn$_4$, Al, and CoSi. The thickness of the three different spin channels was the same as 5 nm, and CoFeB ranged from 2.5 to 12.5 nm with an increment of 2.5 nm. The film thicknesses were estimated by the sputtering rates, calibrated by x-ray reflectivity (XRR) and a contact stylus profiler. For convenience, these three different spin channels were denoted as PtSn, PtSn:Al, and PtSn:CoSi, respectively.

Then, three series of samples, MgO/PtSn$_4$/CoFeB/Al, MgO/PtSn$_4$:Al/Al, and MgO/PtSn$_4$:CoSi/CoFeB/Al were patterned into microstrips of width 5–30 μm and length 70 μm by a photolithography and an Ar ion milling. Then Ti(10 nm)/Au(120 nm) electrical contacts were deposited after the ion milling. Symmetric coplanar waveguides in the ground-signal-ground (CSG) electrodes were used for microwave power injection, which helps avoid the unbalanced perpendicular Oersted field during the spin torque ferromagnetic resonance (ST-FMR) measurements.[21] The resistivities of the thin films were characterized by 4-probe resistance measurements on the microstrips. Keithley's 2182 nanovoltmeter and 6221 current sources were used to probe the voltage and current, respectively. The schematic drawing of the devices, circuits for ST-FMR, and a photograph of a real device were shown in Figure 1.

The crystalline structures of the as-deposited spin channels PtSn, PtSn:Al, and PtSn:CoSi were characterized by x-ray diffraction (XRD) patterns using a Bruker D8 Discover system with a Co $K_\alpha$ radiation point source. All the XRD patterns were converted to a Cu $K_\alpha$ radiation for an easy comparison. The morphologies were characterized by a High-resolution transmission electron microscopy (HRTEM). Elemental mappings were also collected by scanning TEM (STEM) technique using an FEI Titan G2 TEM. The charge-to-spin conversion ratio of the spin channels were characterized by the ST-FMR technique.

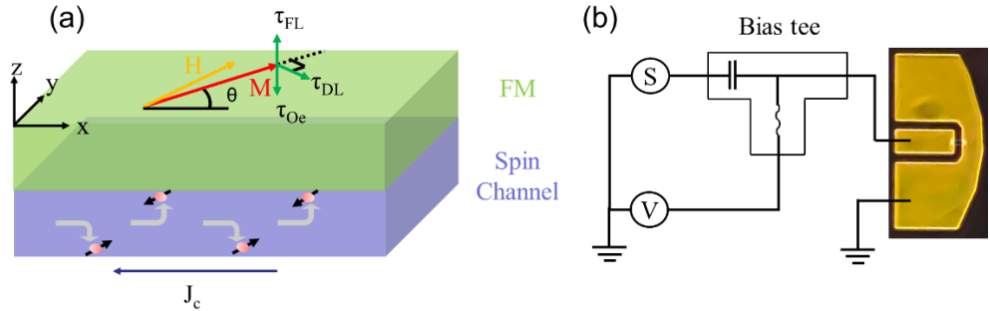

Figure 1. ST-FMR characterizations on the bilayer structure of a spin channel layer and a FM layer. The illustration of spin current generated in the spin channel when a charge current applied and the SOT-driven CoFeB magnetization dynamics in the ST-FMR characterizations (a) schematic of an FM/spin channel bilayer thin film structures. M is the magnetization of the FM layer, and H is the external magnetic field to tune M off its easy axis by an angle of θ. Different torques are then generated as illustrated: field-like torque $\tau_{FL}$, damping-like torque $\tau_{DL}$, and Oersted field torque $\tau_{Oe}$; (b) the schematic circuit used to the ST-FMR characterization. An optical image of an microstrip and GSG electrodes is used to illustrate the setup of ST-FMR measurements.

In the ST-FMR measurement as shown in Figure 1 (a), an AC current with the frequency in gigahertz (GHz) range was applied through the waveguides to a microstrip along its long axis. During the measurement, the frequency of an injected AC current was constant with a magnetic field sweeping at 45 degrees with respect to the long axis of the microstrip. The DC output voltage was recorded at etch magnetic field. As the current flowing into the bilayer structure, damping like torque (DLT) and field like torque (FLT) were generated to the FM layer through the bulk spin Hall effect and interfacial Rashba effect.[2,22–24] A Oersted field torque (OT) was also generated by the charge current in the spin channel layer. The OT in the CoFeB layer canceled out since the thickness of CoFeB is reasonably uniform.

The oscillating spin-orbit torque and Oersted field induce a precession of magnetization in the FM layer, leading to an oscillating resistance due to the AMR in the FM layer expressed as $R(t) = R_0 + \Delta R \cos^2\theta(t)$, where $R_0$ is the resistance when the AC current is perpendicular to the FM layer, $\Delta R$ is the AMR, and $\theta(t)$ is the angle between the AC current and the magnetization of the FM layer. The oscillating resistance is mixed with the oscillating input voltage which gives rise to a DC component and enables the measurements of ST-FMR. A circuit was designed as shown in Figure 1(b) to measure $V_{DC}$. The capacitor allows the AC component to pass through the microstrip, and the inductor lets the DC component be measured by the voltmeter. The DC component fits the Lorentzian formular with the symmetric and antisymmetric contributions. DLT contributes to the symmetric component, while OT and FLT are the origin of the antisymmetric component. The symmetric and antisymmetric components satisfy the Equation (1) below:[25,26]

$$V_{DC} = V_S \frac{\Delta H^2}{\Delta H^2 + (H-H_0)^2} + V_A \frac{\Delta H(H-H_0)}{\Delta H^2 + (H-H_0)^2} \quad (1),$$

where $H$ is the external field, $H_0$ the resonance field, $\Delta H$ the linewidth of the Lorentzian functions, $V_s$ for the strength of the symmetric component, and $V_A$ the strength of the antisymmetric component. The first term in the right part of Equation (1) is the symmetric component from DLT, while the second is the antisymmetric component from FLT and OT.

For the antisymmetric component, OT is related to the thicknesses of the spin channel layer and the FM layer, while FLT is usually determined by the spin channel layer itself, regardless of the thickness of the FM layer.[21] Thus, samples with different thicknesses of FM layer (CoFeB from 2.5 nm to 12.5 nm) were used to separate these two contributions, while the thickness of the spin channel layer was fixed as 5 nm in this work.[26] The ratio of the sum of OT and FLT over DLT can be derived from Equation (1) that separates the contribution of FLT and Oersted field torque : [25]

$$\frac{\tau_{Oe}+\tau_{FL}}{\tau_{DL}} = \frac{V_{Asym}}{V_{Sym}}\sqrt{1+\frac{4\pi M_{eff}}{H_0}} = \left(\frac{J_s}{J_c}\right)^{-1}\frac{e\mu_0 M_s}{\hbar}t_{FM}d_{SC} + \frac{\tau_{FL}}{\tau_{DL}} \qquad (2),$$

where $\tau_{DL}$, $\tau_{Oe}$, $\tau_{FL}$ are damping-like torque, Oersted field torque, and field-like torque, respectively. $t_{FM}$ and $d_{SC}$ are the thickness of FM layer and spin channel layer. And then the equation can be further decomposed to two terms as shown in the right part of Equation (2) to differentiate the contribution of FLT and DLT. In the curve of the torque ratio versus CoFeB thicknesses($t_{FM}$), the slope indicates the charge-to-spin conversion ratio ($J_s/J_c$), and the interception includes the information of FLT. Furthermore, the effective magnetization of FM layer $M_{eff}$ can also be fitted using the Kittel equation that indicates the relationship between the resonance frequency $f$ and resonance field $H_0$:[27] $f = \frac{\gamma}{2\pi}\left[(H_0+H_k)(H_0+H_k+4\pi M_{eff})\right]^{1/2}$, where $\gamma$ is the gyromagnetic ratio, and $H_k$ is the in-plane magnetic anisotropy field.

### 3. Results and discussion

#### 3.1 Morphology and crystal structures

Cross-sectional TEM images of two spin channels, Pt-Sn:Al and Pt-Sn:CoSi with 5 nm CoFeB FM layer and 2 nm Al capping layer, were characterized as shown in Figure 2 (a) and (b). The crystalline structure on the bottom of these images was from the single crystalline MgO substrates. The spin channel layer and the CoFeB layer deposited on the MgO substrates, however, were amorphous as illustrated in the TEM images. The amorphous structure was also demonstrated by the XRD patterns shown in Figure 2 (c), which will be discussed in the next paragraph. The corresponding elemental mapping of these two samples were also characterized. Single elements of Pt, Sn, Al and Co were measured by zooming into the spin channels accordingly. The Si mapping was not included here because a silicon peak can always be detected from the detector itself. In the spin channels, Pt and Sn were uniformly distributed. The dopants were also uniformly dispersed in the spin channels. However, there was a higher dopants concentration region near the interface between the spin channel layer and the CoFeB layer, which might enhance the SOT effects.[28]

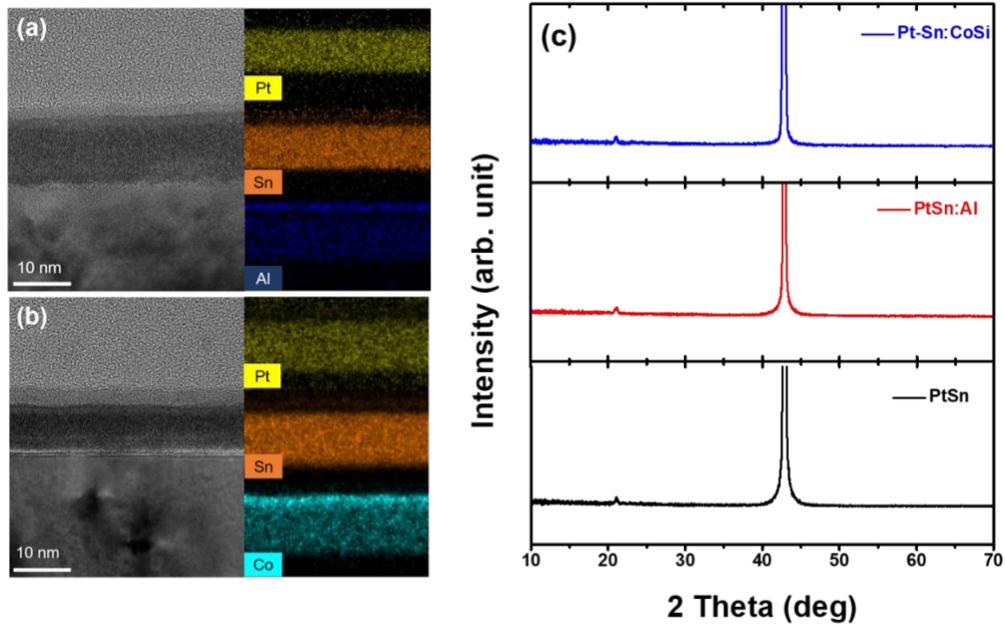

Figure 2. The morphology and elemental mapping of PtSn:Al (a) and PtSn:CoSi (b). The bilayer structure was amorphous for both samples. Elemental mappings indicate dopants dispersed homogeneously in the spin channels, and concentrated at the interface between spin channels and the FM. The XRD patterns of $PtSn_4$, $PtSn_4$:Al, and $PtSn_4$:CoSi only had the diffraction peaks from MgO substrates and no other diffraction peaks were observed, indicating the amorphous nature of the spin channels.

Crystal structures of these three different spin channels PtSn$_4$, PtSn$_4$:Al, and PtSn$_4$:CoSi were characterized using an x-ray diffractometer. The amorphous structure was also confirmed by the XRD pattens of these spin channels as shown in Figure 2(c), which was consistent with the amorphous structure observed in the cross-sectional TEM images. Two diffraction peaks were from the MgO single crystal substrate, where the peak with a lower 2 theta belongs to MgO (100) and the other with a higher 2 theta is MgO (200). However, no other diffraction peaks were observed besides these two from the substrates, indicating all these spin channels have amorphous structures instead of crystalline.

**3.2 Spin torque efficiency**

The spin torque efficiency was characterized by the ST-FMR technique. During the measurement, an AC current with a fixed frequency was injected into a microstrip. Meanwhile, an external magnetic field was sweeping along the direction of 45 degree with respect to the long axis of the microstrip. The DC voltage and the external magnetic field were recorded accordingly at each magnetic field. A resonance peak would occur when the external magnetic field matches to the resonance field of the system. The frequency of input AC currents versus resonance field was plotted in Figure 3 for those three different series of samples: PtSn, PtSn:Al, and PtSn:CoSi with different CoFeB thicknesses ranging from 2.5 nm to 12.5 nm. The data points, denoted by different markers like square, circle, triangle, diamond, and star for different CoFeB thicknesses. These data were fitted by the Kittel equation $f = \frac{\gamma}{2\pi}\sqrt{H_0(H_0 + 4\pi M_{eff})}$ and plotted in the dash-doted lines. The experimental data matched well with the fitting curves, indicating the stable magnetizations in for all the samples.

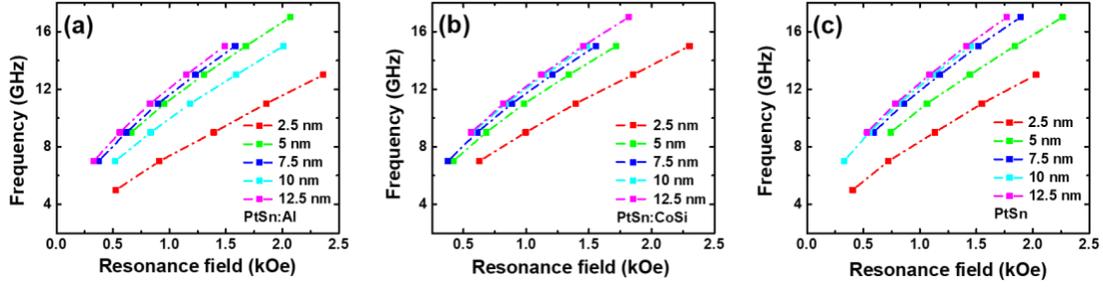

Figure 3. Resonance frequency $f$ versus resonance field $H_0$ in PtSn:Al, PtSn:CoSi, and PtSn spin channels with various CoFeB thicknesses from 2.5 to 12.5 nm. Experimental data of different CoFeB thicknesses were labeled with different markers accordingly, while the dash-dotted lines were the fitting results using the Kittel equation.

For the ST-FMR characterization, an AC current at a certain frequency mixed with the resistance generated voltage $V_{DC}$, depending on the external magnetic field $H$, was plotted in Figure 4 (a), (b), and (c) for PtSn/CoFeB, PtSn:Al/CoFeB, and PtSn:CoSi/CoFeB bilayer samples, respectively. The thickness of CoFeB layer in these samples was 5 nm. A resonant peak appeared when the external magnetic field $H$ equals to the resonant field $H_0$, as illustrated in these curves, under a frequency range of 5 to 17GHz. The ST-FMR data points, denoted by the diamond markers, were fitted using Equation (1), labeled as black dash-dotted lines. The symmetric and antisymmetric components were also fitted and denoted as blue and magenta dash-dotted lines. For the PtSn:Al/CoFeB and PtSn/CoFeB samples, the antisymmetric components had a similar polarity, while the PtSn:CoSi/CoFeB sample was different. This was because the sign of OT and FLT were opposite in the system. At a certain CoFeB thickness, the OT and FLT contributions may cancel each other leading to the zero effective Oersted torque ($\tau_{Oe} + \tau_{FL}$). Thus, the polarity becomes opposite when the CoFeB beyond a critical thickness. The 5 nm CoFeB was close to the transition point of the PtSn:CoSi/CoFeB sample. It turned out that the FLT of the PtSn$_4$ based spin channels studied in this work was not neglectable like some heavy metals and the FLT has significant contributions to the SOT signals.[29,30]

We then plotted the $\frac{\tau_{Oe}+\tau_{FL}}{\tau_{DL}}$ ratio versus the CoFeB thicknesses as shown in Figure 4 (d), (e), and (f), where the black diamond markers indicated the experimental data points, and the blue dashed lines were the fitting results according to Equation (2). The $\frac{\tau_{Oe}+\tau_{FL}}{\tau_{DL}}$ ratio was negative for both PtSn:Al/CoFeB and PtSn/CoFeB samples within the thickness range of CoFeB from 2.5 to 12.5 nm. However, the ratio of PtSn:CoSi/CoFeB sample first was positive 2.5 nm CoFeB and then close to zero when CoFeB became 5 nm and then turned to negative when the thickness of CoFeB exceeded 5 nm. This is also why the antisymmetric fitting curve in Figure 4(b) was so different from the other two because 5 nm CoFeB was close to the critical thickness. Based on the fitting lines, the ratio of $\tau_{FL}/\tau_{DL}$ can also be obtained from the intercept of the dashed lines.

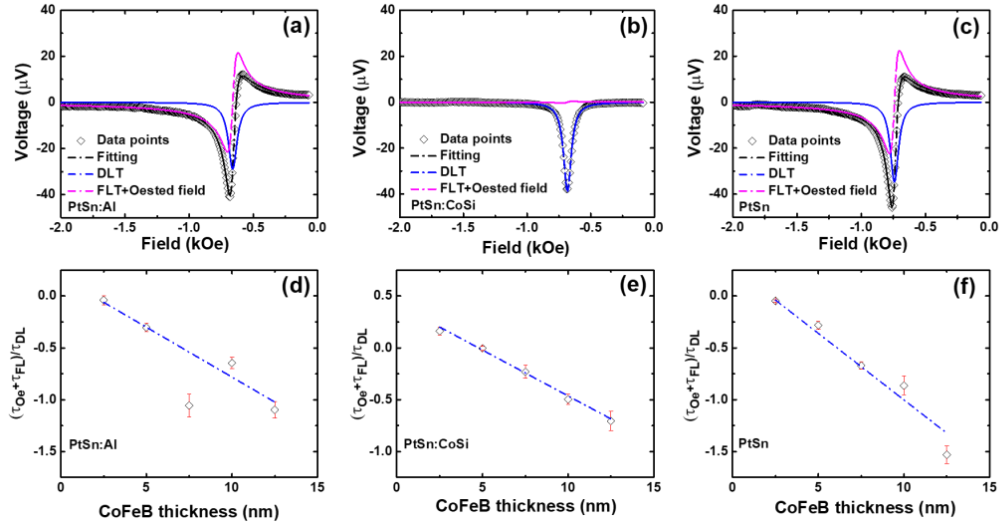

Figure 4. The ST-FMR spectra for the bilayer structure with different spin channels PtSn:Al (a), PtSn:CoSi (b), and PtSn (c) characterized at an excitation frequency of 9 GHz respectively. The CoFeB thickness was 5 nm for all the three samples. The experimental data points were denoted by black diamonds and the fitted symmetric and antisymmetric Lorentzian curves were labeled as blue and magenta with their summation label as black curves. The $(\tau_{Oe} + \tau_{FL})/\tau_{DL}$ ratios for the three series of samples, PtSn:Al (d), PtSn:CoSi (e), PtSn (f), were fitted by Equation

(2). The slops of the fitting lines derive the charge-to-spin conversion ratio, and the intercepts provide the ratio of $\tau_{FL}/\tau_{DL}$.

The charge-to-spin ratios were then calculated based on the slopes of the $(\tau_{Oe} + \tau_{FL})/\tau_{DL}$ vs CoFeB thickness curve according to Equation (2). The FLT and DLT components were plotted in Figure 5 for the three different series of samples accordingly. All the three samples showed similar FLT components, indicating that the amorphous nature of these three types of samples exhibited similar interfacial effects contributing to the FLT. The DLT components of PtSn:Al and PtSn:CoSi, however, had larger charge-to-spin ratio up to 0.14 compared to the PtSn around 0.08. Such enhancement may be due to the Rashba effect at the interface between dopants and the spin channel materials, as Azevedo et al. reported the similar increase of SOC and spin Hall angle in the Pt/YIG bilayer structure by embedding Ag nanoparticles into the Pt layer.[31] The dopants of Al in PtSn$_4$ may also play the similar role like Ag in Pt, which increases the SOC and then the charge-to-spin ratio. It is also possible that the interfacial diffusion between the spin channels and CoFeB layer increase the SOC and contribute to the increased charge-to-spin ratio.[32] Moreover, the resistivity of PtSn$_4$ decreased from 312 μΩ cm to 270 μΩ cm after Al doping, which can further reduce the shunting current in the adjacent FM layer and decrease the critical current needed to switch the FM layer. For CoSi doping, the resistivity increases from 312 μΩ cm to 520 μΩ cm, which indicated the enhanced scattering of carries leading to the increase of charge-to-spin ratio.[33] Even though the CoSi doped samples has a larger coercivity than PtSn$_4$ but still much smaller than topological materials with large resistivities (1000-100000 μΩ cm).[34–37] Lower resistivity topological materials are preferred to minimize the critical current density to switch the adjacent FM layer.[38]

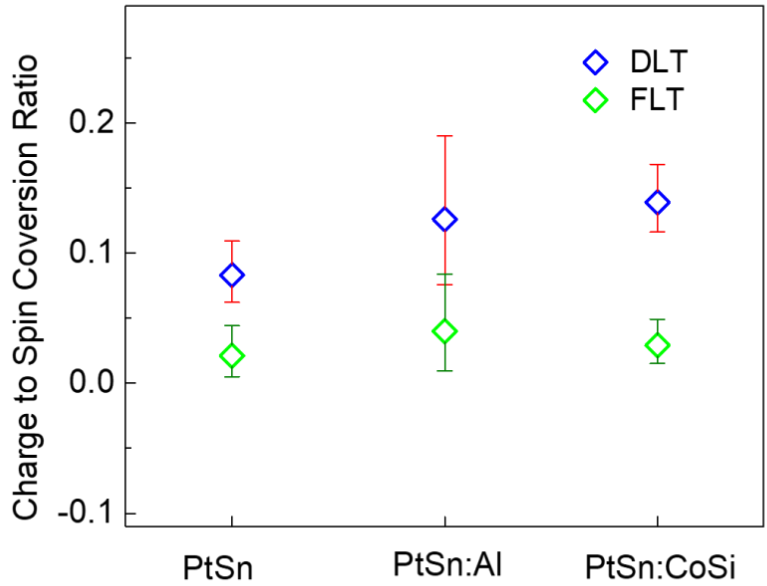

Figure 5. Charge-to-spin ratio of the bilayer structure with different spin channels PtSn, PtSn:Al, and PtSn:CoSi. The mean values were denotated as green circles for the FLT component and blue diamonds for the DLT component.

**4. Conclusions**

In summary, amorphous $PtSn_4$/CoFeB bilayers were synthesized by a magnetron sputtering method. The thickness of $PtSn_4$ layer was 5 nm while the thicknesses of CoFeB ranged from 2.5 to 12.5 nm with an increment of 2.5 nm, which helps differentiate the contribution of DLT and FLT. Two different dopants Al and CoSi were induced in $PtSn_4$ to tune the spin-orbit coupling and enhance the charge-to-spin conversion efficiency from 0.08 to 0.14 of the DLT component. Al doping also decreases the resistivity of the $PtSn_4$ spin channels, which in turn reduces the current shunting in the adjacent FM layer and further decreases the critical current density required to switch the adjacent FM layer.

**Author Information:**

Corresponding Author: * Email: dlzhang@umn.edu (Delin Zhang)

**Author Contributions:**

† J. Liu, Y. Fan, D. Zhang contribute equally to this work

**Acknowledgement:**

This work was supported, in part, by SMART, one of the seven centers of nCORE, a Semiconductor Research Corporation program, sponsored by the National Institute of Standards and Technology (NIST) and by the National Science Foundation through the University of Minnesota MRSEC under Award Number DMR-2011401. Portions of this work were conducted in the Minnesota Nano Center, which is supported by the National Science Foundation through the National Nano Coordinated Infrastructure Network (NNCI) under Award No. ECCS-2025124. D. Zhang and T. Peterson are partially supported by ASCENT, one of six centers of JUMP, a Semiconductor Research Corporation program that is sponsored by MARCO and DARPA. J. -M. L. and J.-P.W. also acknowledge the support from Robert Hartmann Endowed Chair Professorship.